\documentstyle[aas2pp4]{article}

\begin{document}

\title{Strong-field general relativity and quasi-periodic oscillations
in x-ray binaries}

\authoremail{kaaret@astro.columbia.edu}

\author{Philip Kaaret, Eric Ford, and Kaiyou Chen}
\affil{Department of Physics and Columbia Astrophysics Laboratory,
Columbia University, 538 W. 120th Street, New York, NY 10027}

\begin{abstract}

Quasi-periodic oscillations (QPOs) at frequencies near 1000~Hz were
recently discovered in several x-ray binaries containing neutron stars.
Two sources show no correlation between QPO frequency and source count
rate (Berger et al.  1996, Zhang et al.  1996).  We suggest that the QPO
frequency is determined by the Keplerian orbital frequency near the
marginally stable orbit predicted by general relativity in strong
gravitational fields (Muchotrzeb-Czerny 1986, Paczy\'{n}ski 1987,
Kluzniak et al.  1990).  The QPO frequencies observed from 4U~1636-536
imply that the mass of the neutron star is $2.02 \pm 0.12 \, M_{\odot}$.
Interpretation of the 4.1~keV absorption line observed from 4U~1636-536
(Waki et al.  1984) as due to Fe~{\small XXV} ions then implies a
neutron star radius of $9.6 \pm 0.6$~km.

\end{abstract}


\section{Introduction}

Observations of the low-mass x-ray binaries 4U~1608-52 (Berger et al.
1996) and 4U~1636-536 (Zhang et al.  1996) with the {\it Rossi X-Ray
Timing Explorer} (Bradt, Rothschild, \& Swank 1993) have revealed
quasi-periodic oscillations (QPOs) at 800--900~Hz which show no or
little dependence of QPO frequency on source count rate.  In 4U~1608-52,
the QPO frequency changes by less than 10\% as the count rate changes by
a factor of 5 (Berger et al.  1996).  This is in sharp contrast to the
strong correlation of QPO frequency and source count rate seen in the
low-mass x-ray binaries 4U~1728-34 (Strohmayer et al.  1996) and
4U~0614+091 (Ford et al.  1997); in 4U~0614+091, the QPO frequency
changes by more than a factor of 2 as the count rate changes by a factor
of 2 (Ford et al.  1997).  In 4U~1608-52 and 4U~1636-536, the QPO
frequency is, at most, only weakly influenced by the luminosity or mass
accretion rate of the source.  We suggest that the QPO frequency is
determined by the Keplerian orbital frequency near the marginally stable
orbit predicted by general relativity in strong gravitational fields
(Muchotrzeb-Czerny 1986; Paczy\'{n}ski 1987; Kluzniak \& Wagoner 1985;
Kluzniak et al.  1990).

\section{Accretion Flow at the Marginally Stable Orbit}

In general relativity, no stable orbits exist close to a sufficiently
compact massive object.  The radius of the marginally stable orbit
$r_{ms}$ is determined by the mass and rotation of the star and is
independent of the mass accretion rate or radiation flux in the
accretion disk.  In a Schwarzschild spacetime, $r_{ms} = 6GM/c^{2}$,
where $M$ is the mass of the star.  For `soft' nuclear matter equations
of state, the marginally stable orbit can lie outside the neutron star
(Kluzniak \& Wagoner 1985; Cook, Shapiro, \& Teukolsky 1994).  If the
neutron star's magnetic field is sufficiently small, then the accretion
disk will pass through $r_{ms}$.  Keplerian orbits near $r_{ms}$ have
frequencies near 1000~Hz and could generate QPOs which are independent
of source count rate.

An analysis of thin accretion disks by Muchotrzeb-Czerny (Muchotrzeb
1983; Muchotrzeb-Czerny 1986) showed that a stationary flow can not be
supported near $r_{ms}$ for viscosities greater than a critical value.
Stationary flow breaks down at the sonic radius, where the radial
velocity of matter in the accretion disk changes from subsonic to
supersonic.  This instability may lead to observable oscillations either
via Doppler beaming or eclipses and provides a mechanism to generate
QPOs at Keplerian orbital frequencies near $r_{ms}$ (Paczy\'{n}ski 1987;
Kluzniak et al.  1990; Miller, Lamb, \& Psaltis 1996).

The sonic radius is located close to $r_{ms}$, but may move due to
variations in surface density in the inner disk (Muchotrzeb-Czerny
1986).  The analysis of Muchotrzeb-Czerny (1986) suggests that the sonic
radius will oscillate over a relatively narrow range near $r_{ms}$.  If
the surface density variations are irregular, then changes in the QPO
frequency will be irregular.  The QPO frequencies in 4U~1608-52 and
4U~1636-536 show an irregular time evolution, wandering over a small
range in frequency (50~Hz) on a time scale of 1000~s (Berger et al.
1996; Zhang et al.  1996).  The observed range of QPO frequency
variations imply variations in the orbital radius of $\Delta r /r = 2
\Delta \nu / 3 \nu \sim 0.03$.  This is consistent with the allowed
range for sonic radius variations calculated by Muchotrzeb-Czerny (1986)
for $\alpha \sim 0.2-0.3$, where $\alpha$ is a dimensionless viscosity
parameter (Shakura \& Sunyaev 1973).

Motion of the sonic radius should occur on the viscous time scale
$t_{\nu} \sim R^{2}/\nu $ where $R$ is the disk radius and $\nu$ is the
kinematic viscosity.  Using an $\alpha$-prescription for the viscosity,
we find that $t_{\nu} \sim \alpha^{-1} {\cal M}^{2} t_{\phi}$, where
${\cal M}$ is the Mach number of the flow defined as the ratio of the
circular velocity to the sound speed, and $t_{\phi}$ is the orbital time
scale (Pringle 1981).  If the QPOs are associated with Keplerian orbits
then $t_{\phi} \sim 10^{-3}$~s.  For the inner edge of the disk at a
temperature $kT \sim {\rm 1~keV}$, we estimate ${\cal M} \sim 300$.  For
this model to hold, we must have a viscosity greater than the critical
value discussed above, which imples $\alpha > \alpha_{c} \approx 0.03$.
Estimates of $\alpha$ for the inner accretion disk are in the range
$0.1-1$ (Eardley \& Lightman 1975).  Viscosities in this range lead to
viscous time scales of $10^{2}-10^{3}$~s.  This matches the observed
$10^{3}$~s time scale for QPO frequency variations in 4U~1608-52 and
4U~1636-536.

We conclude that motion of the sonic point near the marginally stable
orbit produces small QPO frequency variations which are consistent in
time scale and frequency range with those observed in 4U~1608-52 and
4U~1636-536.  The count rate independence and approximate constancy of
the QPO frequency arises naturally from the count rate independence of
$r_{ms}$.

In one observation of 4U~1636-536, two QPOs are present at frequencies
of $1171 \pm 11$~Hz and $898.7 \pm 1.4$~Hz (van der Klis et al.  1996).
This is similar to the high-frequency QPOs in 4U~1728-34 (Strohmayer et
al.  1996) and 4U~0614+091 (Ford et al.  1996) which have been
interpreted with models in which the higher frequency is the Keplerian
frequency in the inner region of the accretion disk while the lower
frequency `beat' signal results from differential rotation of the inner
disk and the spinning neutron star (Alpar \& Shaham 1985; Lamb et al.
1985).  We suggest that the QPOs near 800--900~Hz in 4U~1608-52 and
4U~1636-536 are beat frequency signals.  Comparison with 4U~1728-34 and
4U~0614-091 supports this identification.  In these two sources the
fractional RMS amplitude of the higher frequency QPO decreases with
count rate while the amplitude of the lower frequency peak increases.
Extrapolation of these trends would lead one to anticipate that only the
beat frequency QPO should be visible in higher luminosity sources such
as 4U~1608-52 and 4U~1636-536.

A possible explanation of why only the beat frequency signal is visible
is that the instability at the sonic point produces a modulation in the
beaming pattern of the radiation but not in the total luminosity of the
source, while a luminosity variation at the beat frequency can result
from radiation feedback from the neutron star onto the disk (Miller,
Lamb, \& Psaltis 1996).  If the optical depth near the neutron star is
large, then the observed amplitude of beaming modulations will be
reduced relative to luminosity modulations (Kylafis \& Phinney 1989),
and, for high optical depths, only the beat-frequency QPO would be
visible.  If the optical depth between the star and the disk is
sufficiently large, the radiation feedback, and therefore the
beat-frequency QPO, may be suppressed (Miller, Lamb, \& Psaltis 1996).

The strong dependence of QPO frequency in 4U~1728-34 and 4U~0614-091 on
source intensity implies that the accretion disks in these sources must
be disrupted before reaching $r_{ms}$.  The qualitiatively different
behavior of the QPOs in 4U~1728-34 and 4U~0614-091 versus those in
4U~1636-536 and 4U~1608-52 is a natural consequence of where the inner
disk is disrupted relative to $r_{ms}$.

\section{Discussion}

If, indeed, the QPOs in 4U~1608-52 and 4U~1636-536 are associated with
Keplerian orbits near the marginally stable orbit, then they provide us
with a tool for determination of the masses of the neutron stars in
these systems (Kluzniak et al.  1990).  For 4U~1636-536, the beat
frequency of 899~Hz, observed simultaneously with the Keplerian
frequency of 1171~Hz, lies near the maximum beat frequency observed.
Since the analysis of Muchotrzeb-Czerny (1986) indicates that the sonic
radius is always equal to or larger than $r_{ms}$ for viscosities above
the critical value, this suggests that the corresponding Keplerian
orbital frequency of 1171~Hz corresponds to an orbit near $r_{ms}$.  The
analysis of Chen and Taam (1993) also shows that the sonic radius lies
very close to, although perhaps inside, $r_{ms}$ for accretion rates
below the Eddington rate.

In a Kerr spacetime, the relation between the mass of the star and the
Keplerian orbital frequency is $M = 2.198 M_{\odot} (\nu_{K}/1000\,{\rm
Hz})^{-1}(1-0.748 j)^{-1}$, to first order in $j$, where $j = c I \omega
/ G M^{2}$ is a dimensionless measure of the angular momentum of the
star (Bardeen, Press, \& Teukolsky 1972; Kluzniak et al. 1990).  The
stellar rotation frequency $\omega / 2 \pi = 272 {\rm Hz}$, and an
assumed moment of inertia $I = 2.0 \times 10^{45} {\, \rm g \, cm^{2}}$,
imply a neutron star mass of $2.02 M_{\odot}$.  Allowing an error of
50~Hz in the Keplerian frequency, sufficient for both the experimental
and theoretical uncertainties, contributes an error of 5\% to the mass
estimate.  Allowing moments of inertia in the range $1-3 \times 10^{45}
{\, \rm g \, cm^{2}}$ contributes an an uncertainty of 4\% to the mass
estimate.  We conclude that the mass of the neutron star in 4U~1636-536
is $2.02 \pm 0.12 M_{\odot}$.  This mass is within the allowed range of
neutron star masses (Kalogera \& Baym 1996) and is not unreasonable
given that accretion onto the neutron star has occured.

It is interesting to note that an absorption feature at $4.1 \pm
0.1$~keV has been observed in thermonuclear x-ray bursts from
4U~1636-536 (Waki et al.  1984).  Similar absorption features have been
seen from 4U~1608-52 (Nakamura, Inoue, \& Tanaka 1988) and EXO~1747-214
(Magnier et al.  1989) by Tenma and EXOSAT.  The most plausible
interpretation of the line-like absorption feature is absorption by
Fe~{\small XXV} ions in matter accreted onto the neutron star surface
after the outburst (Waki et al.  1984; Inoue 1988).  The transverse
Doppler effect is negligible for a neutron star spin rate of 272~Hz.  In
this case, the redshift of the line and the neutron star mass quoted
above implies a neutron star radius of $9.6 \pm 0.6$~km.  This result is
consistent with the mass-radius relation for the AV14+UVII equation of
state (Wiringa, Fiks, \& Fabrocini 1988) and allowing for rotation
(Cook, Shapiro, \& Teukolsky 1994).  We note this equation of state
produces a moment of inertia of $2.0 \times 10^{45} {\, \rm g \,
cm^{2}}$ for a $2.0 M_{\odot}$ star, consistent with our assumed $I$
above.

If confirmed, the relation between high-frequency QPOs in x-ray binaries
and the marginally stable orbit in general relativity may open a new
experimental venue for the study of strong gravitational fields and will
permit the determination of neutron star masses as described above.
Further work is required to improve our theoretical understanding of
accretion flows near $r_{ms}$ and to obtain more detailed observations
of high-frequency QPOs and the x-ray binary systems that produce QPOs.
Independent determination of the mass of the neutron star in
4U~1636-536, via measurement of the binary orbital parameters, would be
of great interest.  Comparison of the mass with the radius of the
marginally stable orbit determined from the QPO frequency would permit a
test of strong field and inertial frame dragging effects in general
relativity (Kluzniak et al. 1990).

\acknowledgements
We thank M.C. Miller, F.K. Lamb, and M. Ruderman for useful discussions.


\end{document}